# Dissociative Electron Attachment Cross Sections for $Ni(CO)_4$, $Co(CO)_3NO$, $Cr(CO)_6$

[1]Maria Pintea, [1]Nigel Mason, [2]Maria Tudorovskaya

[1]School of Physical Sciences, University of Kent, CT2 7NZ, Canterbury, UK

[2]Quantemol Ltd, EC1V 2NZ, London

**Abstract**

The $Ni(CO)_4$, $Cr(CO)_6$, $Co(CO)_3NO$ are some of the most common precursors used for focused electron induced deposition. Some of the compounds, even though extensively used have high requirements when it comes to handling them being, explosives, highly flammable and with high toxicity level, as is the case of $Ni(CO)_4$. We are employing simulations to determine values that are hard to determine experimentally, and compare them with DFT calculations and experimental data where available. 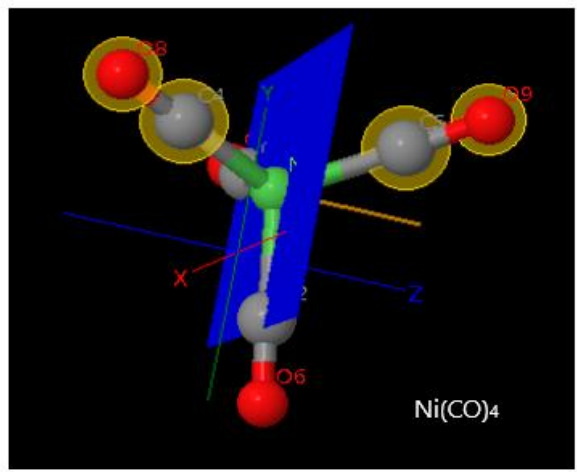 Using Quantemol-N cross section simulations for dissociative electron attachment (DEA) at low electron energy, 0 – 20eV, gives valuable information on the fragmentation of the molecules, using their bond dissociation energies, electron affinities and incident electron energies. The values obtained for the cross sections are $0.12 \times 10^{-18} cm^2$ for $Ni(CO)_4$, $4.5 \times 10^{-16} cm^2$ for $Co(CO)_3NO$ DEA cross-sections and $4.3 \times 10^{-15} cm^2$ for $Cr(CO)_6$.

**Introduction**

As Focused Electron Beam Induced Deposition[1] is developing with the possibility of becoming a viable manufacturing technique, the need to have more data available on molecules of interest, alkenes, silanes, metal halogens, carbonyls, phosphines, acetylacetonates, [1] increases. The limitation of this direct - write fabrication technique comes in the appearance of secondary and backscattered electrons as part of the primary electron beam or secondary electron beam. The effect of the secondary and backscattered electrons at low electron energy level, 0 – 50eV, is the deposition of a thin halo and creation of secondary structures in the vicinity of the primary structures as well as incomplete fragmentation of the precursors reducing the purity of the final structure. [2] To analyze and recreate this effects, gas phase and surface science studies are employed. In gas phase studies, the interaction of molecules with single electrons is evaluated and resulting fragmentation pathways analyzed. On the other hand, in the ultra-high vacuum surface science setups, the interaction of the electrons with the molecules and the substrates is evaluated, identifying the species desorbing from the substrate. However for many of these FEBID compounds, an increase in such information as the probability of collision between electrons and molecules, and the dynamics of the processes underlying the induced chemistry in organometallic compounds is needed and cross sections for: dissociative ionization, elastic scattering, vibrational excitation, dissociative electron attachment, neutral dissociation, bipolar dissociation. The particular metal containing compounds that will be used as the base of our simulations for cross sections determination are compounds that have been extensively researched and used for the deposition purposes and commercially available, but scarce cross section data has been published on them. The data presented from our R-matrix calculations will focus on four widely used compounds $Ni(CO)_4$, $Co(CO)_3NO$ and $Cr(CO)_6$ commercially available for purchase and widely used in Focused Electron Beam Induced Deposition.

**Molecular complexes used for calculations**

The carbonyl group compounds ($Ni(CO)_4$, $Cr(CO)_6$ and $Co(CO)_3NO$) are relatively easy to purchase commercially simple symmetric structures make them suitable for electron-induced chemistry applications and potentially suitable for creating very clean deposits in the focused electron beam induced process with purity over 90% and low resistivity. The structural representation of the three compounds is presented in Fig 2 and the X, Y, Z coordinates used for Quantemol-N calculations are presented in ANNEX 1.

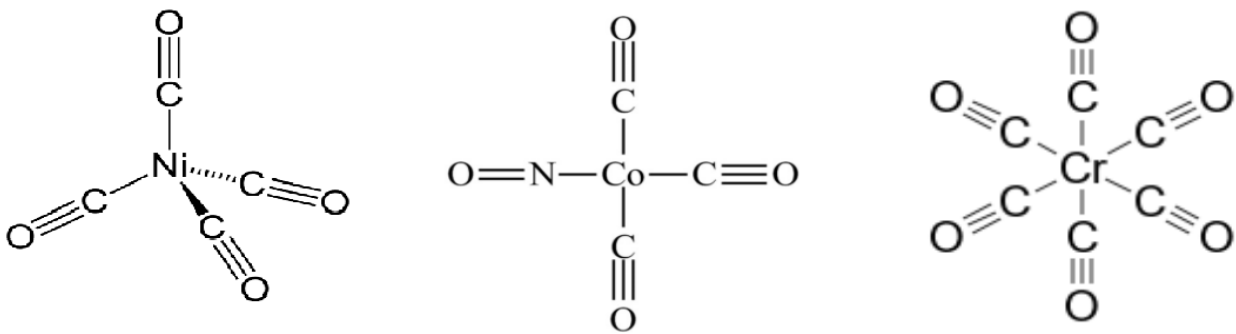

Fig 2. Molecular complexes: $Ni(CO)_4$, $Co(CO)_3NO$ and $Cr(CO)_6$

The Ni(CO)$_4$ is less used in the industry due to its high toxicity, similarly to Cr(CO)$_6$; it is highly flammable and insoluble in water; its lowest decomposition temperature is 50°C, making it hard to be used at room temperature and requiring cryogenic conditions. The Cr(CO)$_6$ compared to the Ni(CO)$_4$ has higher volatility and is more stable at room temperature with a thermal decomposition point over 150°C. Similar to the Cr(CO)$_6$, the Co(CO)$_3$NO has a high decomposition temperature, 130 – 140°C. Detailed Raman, FTIR and MOs studies are compared in Cross-section section with the theoretical model obtained from our calculations.

**R-matrix theory and method of calculation**

Scattering The scattering calculations based on the R-matrix [13] theory are carried out to determine the cross-sections of electron scattering by Ni(CO)$_4$, Cr(CO)$_6$ and Co(CO)$_3$NO. We use Quantemol-N simulation package interfacing UKRmol code suite [52]. Quantemol-N can be used to compute a number of cross-sections for elastic and inelastic electron scattering, the most important to the induced chemistry being the dissociative electron attachment (DEA) cross-sections[15].

The DEA process is a process widely found in nature due to the formation of negative ions at relatively low electron energies, though our applications limit themselves to focused electron beam induced processes. The dissociation process and the induced chemical reactions follow the schematics:

$$AB + e^- \rightarrow AB^* \rightarrow A^- + B \text{ or } A + B^- \tag{1.1}$$

which can be followed by the dissociation reaction:

$$M(CO)_n \rightarrow M(CO)_{n-m} + m(CO), \text{ where } M = Cr, Ni, Co \tag{1.2}$$

The DEA process can undertake two paths in the evolution of the compound from gas - phase molecules to fragments or radicals: the first path (a) is the irradiation of the molecule with an electron that would cause a change in the molecule's energy found in a certain resonant state, undergoing a jump to a higher ionized state, and the second path (b) is either a repeated autoionization or if the resonance state has a longer lifetime, dissociation of the molecule into fragments.

The total dissociative attachment cross section is is a weighted sum of DEA cross-sections, $\sigma_i$, over all resonances in the scattering process (1.3):

$$\sigma_T(E) = C \, \Sigma_i \, S_i \, \sigma_i(E) \tag{1.3}$$

The positions of the resonances found with the UKRmol routine RESON [53]. S is the survival probability for the resonance or the probability of a resonance, E is the incident electron energy, and C is the adjustment factor[15]. It is important to note that in the model used by Quantemol-N, the fragments are moving in the effective potential and therefore behave like a quasi-diatomic molecule. For each resonance, the partial cross-section has the Breit-Wigner shape:

$$\sigma_{BW}(E, r) = (2\pi/E) (\Gamma^2/4) / [(E-V(r))^2 (\Gamma^2/4)] \quad (1.7)$$

where $\Gamma$ is the width of the resonance, r is the distance between the dissociating fragments, and V(r) is the effective potential.

The resonances positions and widths are determined by the UKRmol codes which treat the incoming electron in the same way as the molecule's electrons in its vicinity inside of the so-called R-matrix sphere. The scattering wavefunction can be expanded over the target states and expressed in terms of N-electron target wavefunctions, continuum orbitals representing the scattering electron inside the R-matrix sphere, and additional quadratic integrable functions constructed from the target occupied and virtual molecular orbitals [13]. Far away from the molecule, the scattered electron moves in the effective potential.

Whilst the *ab initio* part of the calculation is quite rigorous, further assumptions about the dissociation channels and assigning resonances to a specific channel may introduce uncertainty.

The cross section values from Quantemol-N software are in very good agreement with the experimental data available for the particular molecules. In preparing the present calculations we benchmarked the current calculations producing a set of $CH_4$ cross section data, we obtained the same results for the total cross section and inelastic cross sections, which in previous Quantemol calculations replicated experimental data.

**Cross Sections**

**Co(CO)$_3$NO.** The Co(CO)$_3$NO compound has $C_{3v}$ symmetry, with the three CO groups on the faces of a tetrahedron and the NO group to Co(IV). The total cross section is high with values of $1.4 \times 10^{-12}$ cm$^2$ for an energy range of 0 - 100eV. Engmann et al (2013) measured the DEA cross section estimating the maximum DEA cross section having a value of $4.1 \times 10^{-16}$ cm$^2$ for the loss of only one CO ligand this being the predominant DEA process and the DI cross sections with values of $4.6 \times 10^{-16}$ cm$^2$.

| Compound | Dissociation | BDE (kcal/mol) |
|---|---|---|
| $Co(CO)_3NO$ | $Co + 3CO + NO$ | 144.8 – 154.4[16] |
| $Ni(CO)_4$ | $Ni(CO)_3 + CO$ | 35[21], 22.3[44] |
| $Co(CO)_3NO$ | $Co(CO)+2CO+NO$ | 115[16] |
| $Cr(CO)_6$ | $Cr(CO)_5 + CO$ | 49.8[26], 38[25] |

Table 1. Bond dissociation energies for $Co(CO)_3NO$, $Ni(CO)_4$, $W(CO)_6$

The spectroscopy of $Co(CO)_3N^{15}O$ and $Co(CO)_3N^{14}O$ in vapor form has been analyzed [6], with the infrared spectrum giving vibrational band frequencies at 2108cm$^{-1}$, 2047cm$^{-1}$, 1822cm$^{-1}$ for C - O and N - O stretch and 2010cm$^{-1}$ for $C^{13}$ - O isotopic species of $Co(CO)_2(C^{13}O)NO$. Bartz et al (1998) determined the highest excited state of $Co(CO)_3NO$ y2F5/2 at 36300cm$^{-1}$ equivalent to 103.8 kcal/mol, the three Co - CO bonds and one Co - NO bond needing an extra energy of 154.4 kcal/mol to break the ligands and 38 kcal/mol as the adiabatic metal - ligand bond dissociation energy for two CO groups.

At higher energy, the study of Rosenberg et al (2013) splits the problem in two parts, for irradiation with electron densities less than 5 x 10$^{16}$ e$^-$/cm$^2$ where the C(1s) peaks appear at 287.8eV and 293.3eV, a π-π* transition, the N(1s) that has the peak at 401.6eV, the O(1s) peaks at 534.0eV and 534.6eV and an oxide peak at 529.7eV, and the Co(2p$_{3/2}$) peak at 780.9eV. For electron densities over 5 x 10$^{16}$ e$^-$/cm$^2$, the C(1s) peak appears only at 285eV, the N(1s) peak does not change its position, the O(1s) peaks attenuate in amplitude and the oxide peak at 529.7eV now increases in amplitude. If the concentration ratios of after and before irradiation are taken into account, a decrease in the values of $O/O_0$ and $C/C_0$ is observed and $N/N_0$ remains constant. The same study presents a reflection absorption infrared spectroscopy study (RAIRs) of the Co - CO and Co - NO bonds, as appearing at 2111cm$^{-1}$ and 2064cm$^{-1}$, and respectively 1827cm$^{-1}$. Comparatively different studies present these vibrational frequencies at 2108cm$^{-1}$ and 2047cm$^{-1}$, and NO stretch frequencies at 1822cm$^{-1}$ (McDowell et al (1961)), and 2100.3cm$^{-1}$ and 2033cm$^{-1}$, respectively Co - NO vibrational frequencies at 1806.2cm$^{-1}$ (Horrocks et al (1963)).

Sawyer et al (2008) determine experimentally the vibrational frequencies at photoinduced dissociation through DFT calculations and picosecond time-resolved infrared spectroscopy showing two bands between 200nm and 400nm. Two types of possible geometries are analyzed, the linear Co – N - O and 120 - 170° angle geometry for the Co - NO bond, giving the vibrational frequency bands for Co - NO at 1808cm$^{-1}$ in close agreement with the experimental value of 1842cm$^{-1}$, 1684cm$^{-1}$ and 1715cm$^{-1}$. The CO stretching mode appears at 2037cm$^{-1}$ and 1980cm$^{-1}$ [17]. The presence of the same bent Co – N - O excited state is demonstrated by the anisotropic angular distribution [18].

The DEA process is presented in showing the sequential nature of the fragmentation (1.9):

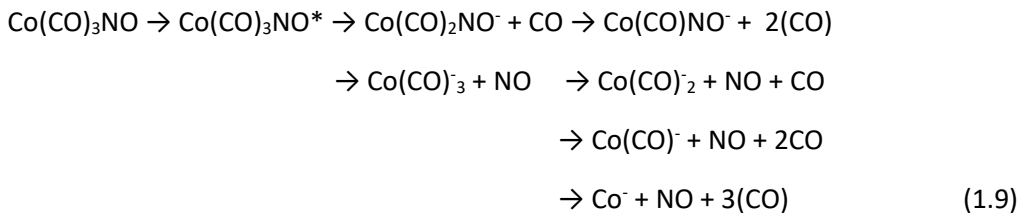

$Co(CO)_3NO \rightarrow Co(CO)_3NO^* \rightarrow Co(CO)_2NO^- + CO \rightarrow Co(CO)NO^- + 2(CO)$

$\rightarrow Co(CO)^-_3 + NO \quad \rightarrow Co(CO)^-_2 + NO + CO$

$\rightarrow Co(CO)^- + NO + 2CO$

$\rightarrow Co^- + NO + 3(CO) \qquad (1.9)$

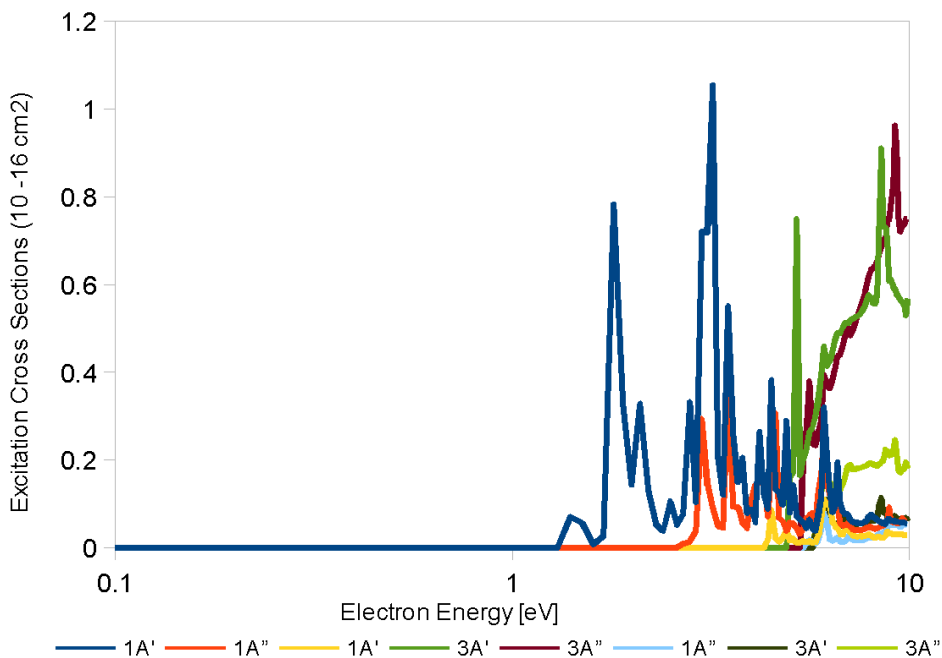

Fig 3. Inelastic Scattering Cross Sections for Co(CO)$_3$NO from Quantemol-N simulations

Knowing that the dissociative electron attachment process happens in the range of 0-15eV (see Table 2), the negative ions are formed at incident electron energies of 0.2eV to 7.1eV, all relatively low electron energies with values under 10eV. The formation of pure Co$^-$ takes place at 7.1eV, stripping off a (CO)$_3$NO radical. The dissociative electron attachment to Co(CO)$_3$NO for the dissociation of one CO ligand at 0.9eV with the formation of Co(CO)$_2$NO$^-$ ion is presented in Fig. 3 with values of ~22.5Å. The incident bond dissociation energy for the formation of all negative ions used in our calculations is presented in Table 2.

The Co(CO)$_3$NO has 84 electrons and it's ground state is a closed shell $^1A_1$ state with C$_{3v}$ symmetry and C$_1$ work point group with its lowest lying excited state $^3$A. The Co - NO is at 160.8° and Co – N - O is 4A" state at 149.5°. With the two bond dissociation energies, for Co - NO bond at a value of 1.63eV and for the Co – CO bond at 1.26eV [45], the initial ground state in the fragmentation of Co(CO)$_3$NO is $^3$A" with the excitation states $^1$A', $^1$A", $^1$A', $^3$A', $^3$A", $^1$A", $^3$A', and the $^3$A" giving the symmetry scattering resonances from the triplet state as ²A', ²A", ⁴A' and ⁴A". To simplify our Quantemol-N simulations, a $C_{2h}$ geometry of the molecule was used. The active space from the Quantmeol-N simulation is 29A', 30A', 13A'', 14A''.

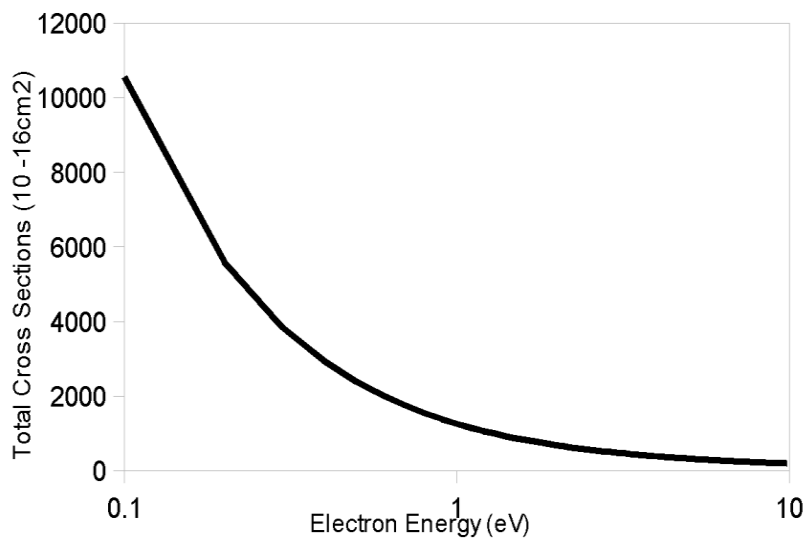

Fig 4. Co(CO)₃NO total cross sections from Quantemol-N

The ionization energy for the Co(CO)₃NO has a value of 8.33eV for an electron affinity of 0.75eV, with the ionization threshold at a value of 9.2eV. The presented reaction energy for the dissociation of the Co(CO)₃NO into Co(CO)₂NO⁻ + CO is in the range of 0.65eV.

| Negative Ion | Incident Electron Energy (eV) [7] | Ion Formation | Incident Electron Energy (eV) [41] | Negative Ion | Incident Electron Energy (eV) [32] |
|---|---|---|---|---|---|
| Co(CO)₂NO⁻ | 0.9 | Ni(CO)₃⁻ | 0.8 | Cr(CO)₅⁻ | 0.1 |
| CoCONO⁻ | 2 | Ni(CO)₂⁻ | 1.7 | Cr(CO)₄⁻ | 1.5 |
| CoNO⁻ | 5 | Ni(CO)⁻ | 4.6 | Cr(CO)₃⁻ | 4.7 |
| Co(CO)⁻₃ | 1.8 | Ni⁻ | 5.4 | Cr(CO)₂⁻ | 5.9 |
| Co(CO)⁻₂ | 3 | | | CrCO⁻ | 8.5 |
| CoCO⁻ | 6.4 | | | Cr⁻ | 8.8 |
| Co⁻ | 7.1 | | | | |

Table 2. Negative Ion Formation for Co(CO)₃NO, Ni(CO)₄ and Cr(CO)₆

The dissociative electron attachment cross section for Co(CO)₃NO is presented according to the electron affinity, bond dissociation energy and incident electron energy. The electron affinity of the negative ions formed through the dissociative electron attachment process are discussed in [27] with values close to

1.35eV < EA[Co(CO)$_2^-$] < EA[Co(CO)(NO)$^-$] < EA[Co(CO)$_3^-$] < EA[Co(CO)$_2$NO$^-$] = 1.73eV. The simulations were run taking into account these differences.

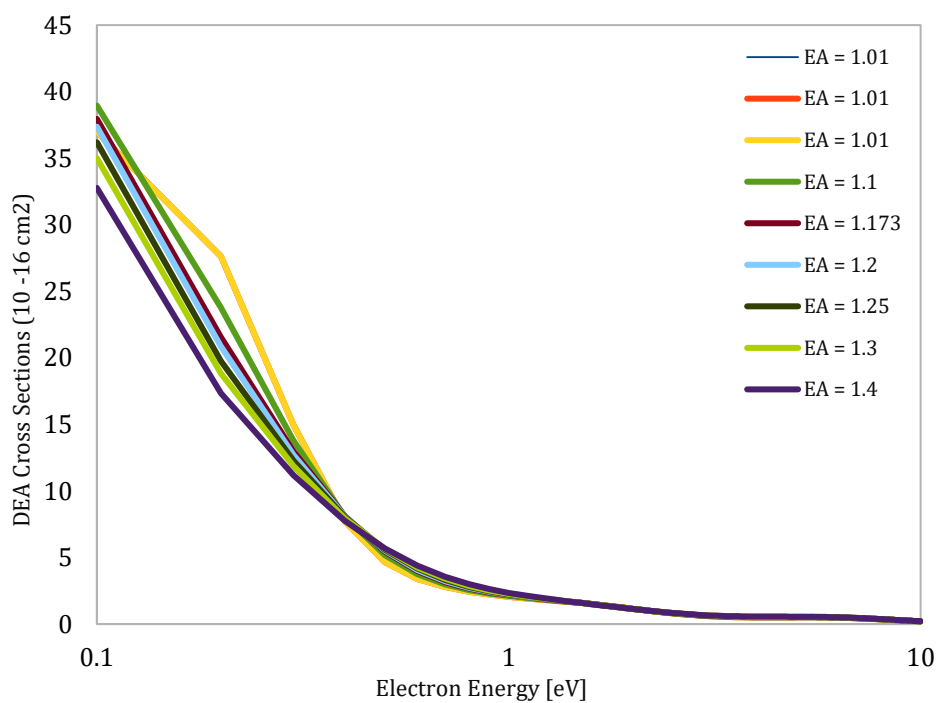

Fig 5. Dissociative electron attachment cross sections from Co(CO)$_3$NO

The maximum of the cross section is in the range of 4.5 x 10$^{-16}$cm$^2$ at 0.1eV corresponding to the Co(CO)$_2$NO$^-$ fragment.. The individual peaks in the DEA cross section graph are corresponding to the individual fragment dissociation, close in energy to the eV value we found in the literature, though the individual cross section values are very low compared to the computed ~0eV value. The maximum cross section value changes with the vibrational frequency. The purple (color) (Fig. 3) curve is close to reality, presenting a clean peak at a vibrational frequency of 0.53Å$^2$ and 1.173eV electron affinity.

**Cr(CO)$_6$.** The Cr(CO)$_6$ bond distances we used for our simulations are 1.916Å between Cr – C and 1.171Å between C – O and the symmetry point group of the ground state molecules is O$_h$, though different sets of values have been reported in [50]. The bond distances reported in Cr(CO)$_6$ are 1.926Å for in axis Cr – C and 1.139Å for C – O. The values are reduced for the equatorial bonds for Cr – C with a value of 1.918Å and C – O with a value of 1.141Å.[50] This values have been used for the structure of the Cr(CO)$_6$ in our Quantemol-N calculations. The symmetry point group of the molecule is O$_h$, but for our simulations simplification we used a D$_{2h}$ configuration.

| Molecule | Bond distances (Å) CCSD (Cr – C) [51] | Bond distances (Å) CCSD (C – O) [51] | Bond distances (Å) CCSD (Cr – C)$_{eq}$ [50] | Bond distances (Å) CCSD (C – O)$_{eq}$ [50] |
|---|---|---|---|---|
| Cr(CO)$_6$ | 3.684 | 2.207 | 1.918 | 1.141 |

Table 3. Bond distances for Cr(CO)$_6$

The structural and symmetry data used is presented in Annex 1. Whitaker and Jefferey [28] determined the space group of Cr(CO)$_6$ as the *Pnma* or *Pn2$_1$a*. The dissociation process in Cr(CO)$_6$ is following the steps:

$$Cr(CO)_6 + e- \rightarrow Cr(CO)_6 \rightarrow Cr(CO)^-_x + n(CO), \text{ where } n = 6 - x \qquad (1.10)$$

The dissociation of Cr(CO)$_6$ into Cr(CO)$_5^-$ and a (CO) radical at 0.1eV, as a result of the dissociative electron attachment (DEA) process, is a transition from the lowest lying LUMO orbital 9a$_{1g}$ (σ) or one of the higher lying virtual orbitals (π) to a HOMO higher energy state orbital (σ*) or to the highest unoccupied HOMO orbital (π*).

The optical spectra of Cr(CO)$_6$ presents the only one allowed spin transitions $^1A_{1g} \rightarrow {}^1T_{1u}$ as well as multiple other smaller bands assigned to $^1A_{1g}(2t^6_{2g}) \rightarrow {}^1T_{1g}$, $^1T_{2g}(2t^5_{2g}6e^1_g)$ transitions. [29], [35] The 3.5eV to 7eV was assigned [29] to $^1A_{1g} \rightarrow {}^1T_{1u}$, while the 4.83eV was assigned to $^1T_{2g}$ and 4.91eV to $^1A_{1g} \rightarrow a^1T_{2g}$, 3.60eV and 3.91eV was assigned to the allowed transition $^1A_{1g} \rightarrow a^3T_{1u}$. Our ions fall in the energy range between 0.1eV to 8.8eV.

| Compound | Fukuda et al [36] | Winters and Kiser [37] | Foffani et al [38] | Electron Ionization [11] | Photon Impact [11] | Metal Atom [11] | Junk and Svec [11] |
|---|---|---|---|---|---|---|---|
| Cr(CO)$_6$ | 8.5 | 8.15 ± 0.17 | 8.18 ± 0.07 | 8.23 | 8.03 | 6.76 | 8.44 ± 0.05 |

Table 3. Ionization potential of Cr(CO)$_6$

Villaume et al [31] defines the structure of the molecule as D$_{2h}$ and the electronic ground state $^1A_{1g}$ as having the electronic configuration $(8a_{1g})^2(7t_{1u})^6(1t_{2g})^6(1t_{2u})^6(1t_{1g})^6(5e_g)^4(8t_{1u})^6(2t_{2g})^6$. At 4.48eV and 4.50eV, the allowed transition is a $a^1A_{1g} \rightarrow {}^1T_{2u}$. At a temperature of 77K, vibrational bands appear at 4.05eV, 4.49eV; at 300K the vibrational bands can be found at 5.53eV, 4.87eV and 6.36eV, all assigned to $a^1A_{1g} \rightarrow {}^1T_{1g}$ and $a^1A_{1g} \rightarrow {}^1T_{2g}$.

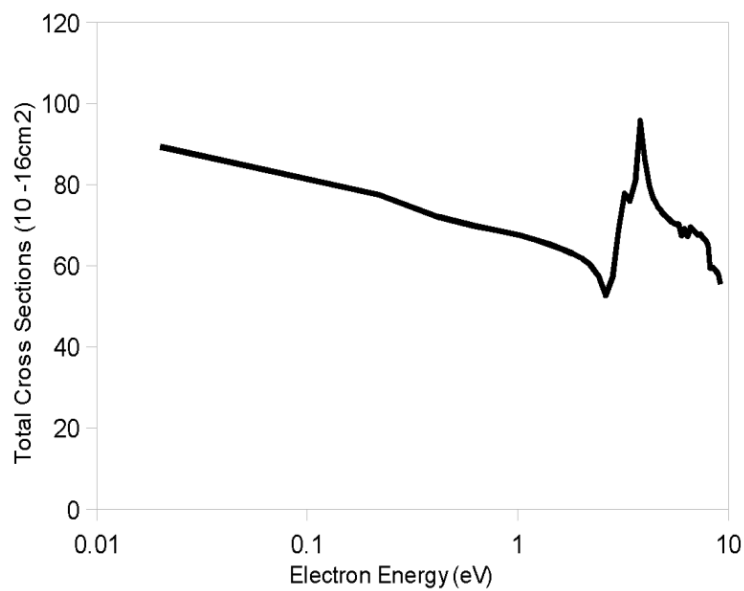

Fig 6. Cr(CO)$_6$ total cross sections from Quantemol-N

| Negative Ion | Dissociation mechanism (eV) from Cr(CO)$_6$* | Experimental Incident Electron Energy (eV) [32] | Electron Affinity (eV) | Vibrational Frequency (cm$^{-1}$) |
|---|---|---|---|---|
| Cr(CO)$_5^-$ | Cr(CO)$^-_5$ + CO | 0.1 | >1.6eV [38] | 2000 |
| Cr(CO)$_4^-$ | Cr(CO)$^-_4$ + 2(CO) | 1.5 | | |
| Cr(CO)$_3^-$ | Cr(CO)$^-_3$ + 3(CO) | 4.7 | | |
| Cr(CO)$_2^-$ | Cr(CO)$^-_2$ + 4(CO) | 5.9 | | |
| CrCO$^-$ | Cr(CO)$^-$ + 5(CO) | 8.5 | | |
| Cr$^-$ | Cr$^-$ + 6(CO) | 8.8 | | |

Table 4. Negative ions of Cr(CO)$_6$ with electron affinity

The ionization potential for the ions of the Cr(CO)$_6$ molecule have values higher than the threshold value of ~8eV. Table 3 presents the ionization potential of the parent Cr(CO)$_6$ molecule from the publications of Winters and Kiser [37], Fukuda et at [36], Foffani et al [38] showing a ionization potential for carbonyls with a value of 1.5eV higher than for the metal atom and similar values within 0.1 – 0.5eV.

The basis set used for our Quantemol-N calculations on cross sections are cc-pVDZ for C atom, 6-311G for O atom and cc-pVTZ on Cr .

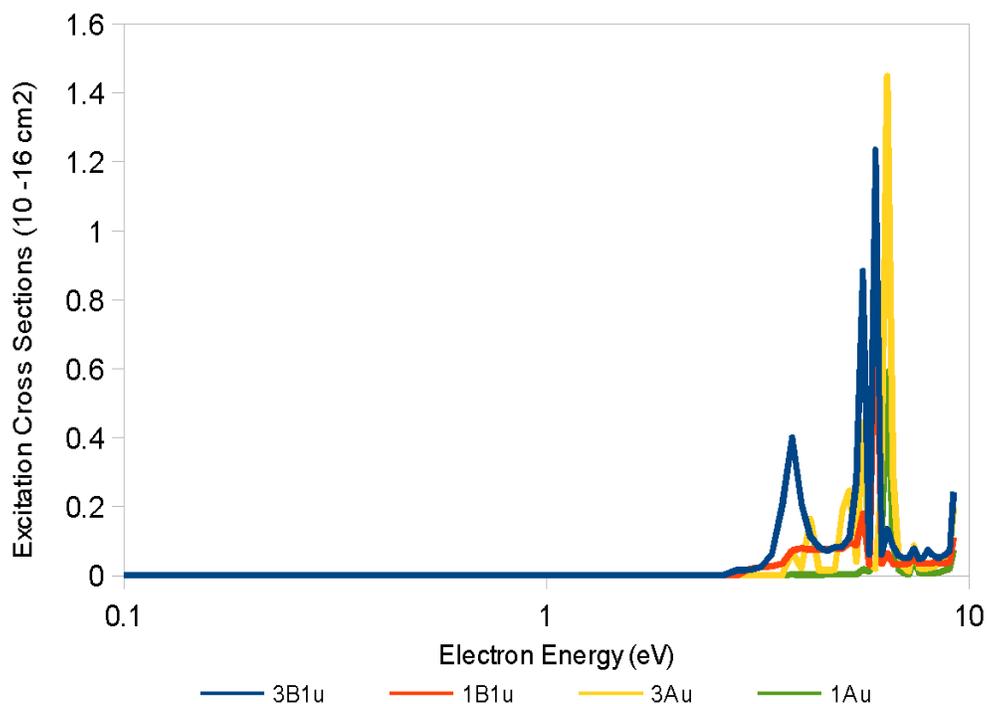

Fig 7. Excitation cross sections of Cr(CO)$_6$

The DFT calculations in [33] reveal the vibrational excitation bands due to the photoionization of the molecule and show the process as being a transition from de$_g$* orbitals in σ-antibonding with the metal and a reduction in the number of electrons in π-bonding t$_{2g}$ orbitals to occupying the Cr-d$z^2$ – CO-5σ e$_g$*orbitals. The cross section data was calculated using Quantemol - N simulations using dissociation electron energy data and electron affinity of the formed negative ions presented in Table 4.

| State | TDDFT/B3LYP - ΔE(eV)[50] | Experimental [50] ΔE(eV) | Experimental [32] /Our Q-N* Sim. ΔE(eV) |
|---|---|---|---|
| 1 $^1E_u$ | 4.14 | | |
| $1^1T_{2u}$ | 4.2 | | |
| $1^1A_{2u}$ | 4.25 | | |
| **$1^1T_{1u}$** | **4.5** | **4.44** | **4.7** |
| $1^1T_{1g}$ | 4.65 | | |
| $1^1A_{1u}$ | 4.7 | | |
| $2^1T_{2u}$ | 4.82 | | |
| $2^1E_u$ | 4.71 | | |
| $1^1E_g$ | 4.99 | | |
| $1^1T_{2g}$ | 4.74 | | |
| $2^1T_{1g}$ | 4.91 | | |
| $2^1T_{2g}$ | 5.59 | | |
| $2^1T_{1u}$ | **6.02** | **5.48** | **5.9** |

Table 5. Vibrational excitation energies of $Cr(CO)_6$; (Q-N* used for Quantemol-N simulations)

If the $Cr(CO)_6$ molecule is seen in $C_{2v}$ point group [50], the allowed transition from $1^1A_1$ happens to $T_{2u}$ state that is characterized by 3 sub-states: $1^1B_1$, $1^1B_2$ and $2^1A_1$, where $1^1B_1$ and $1^1B_2$ are degenerate states corresponding to $1^1E$ in a $C_{4v}$ point group. A Cr – C bond distance higher than 0.25Å the energy levels of the transition states are: $E(1^1B_1) = E(1^1B_2) > E(1^1A_2)$ for $C_{2v}$ point group and $E(1^1E) < E(1^1B_2)$ for $C_{4v}$ point group. The vibrational and transition states data is presented in Table 5.

The DEA cross sections are presented in Fig. 3, simulated with different electron affinity for comparison in the spectrum changes. The values used for the electron affinity of the negative ions of $Cr(CO)_6$ are expected to be higher than the specified value in [38] of 1.6eV. The DEA cross section maximum value is in the range of ~4.3 x $10^{-15}$cm² at 0.2ev for the $Cr(CO)_5^-$ fragment, for a value of electron affinity of 1.6 eV and vibrational frequency of 2000 cm$^{-1}$ for a dissociation electron energy value of 0.1eV. The value of cross-sections reported by [50] is of 1.85Å² to 3.29Å².

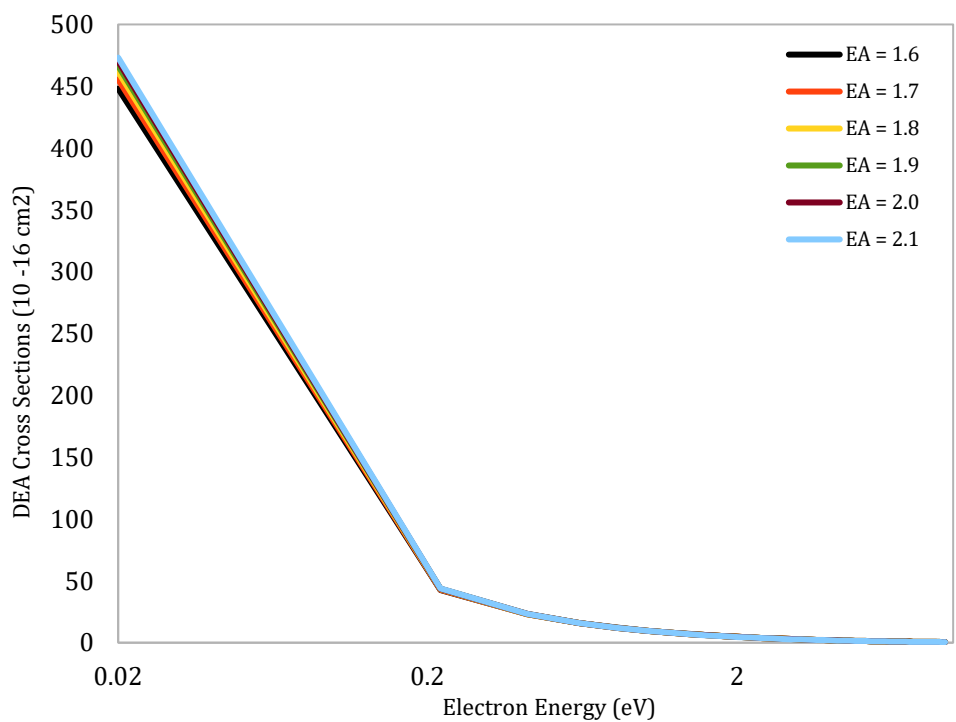

Fig 8. DEA cross sections of $Cr(CO)_6$

The ground state configuration used in our Quantemol-N calculations in $^1A_1$ state is $D_{2h}$. The final assigned states for the DEA process as a result of the calculations from the initial $^1A_g$ state are: $^3B_{1u}$, $^1B_{1u}$, $^3A_u$, $^1A_u$ with 5 symmetry states. From Quantemol-N simulations we get the resonances from our active space are: $^2A_g$ (4.86eV), $^2B_{3u}$ (9.04eV), $^2B_{2u}$ (9.04eV), $^2B_{1g}$ (8.1eV), $^2B_{1u}$ (8.1eV), $^2B_{2g}$ (8.3eV), $^2B_{3g}$ (8.3eV) and $^2A_u$ (6.7eV).

**Ni(CO)$_4$.** The Ni(CO)$_4$ is a tetrahedral, $D_h$, with its ground state $^1A_1$. The Ni - n(CO) bond is a d → π* transition, from the lower stable orbitals d orbitals with ground state ($^1A_1$) to the higher π* orbitals of t-symmetry with the ($1^1T_1$, $1^1T_1$, $3^1T_1$, $1^1E$, $2^1T_2$) excited states. [20] The UV spectrum of Ni(CO)$_4$ in [30] has peaks at 6.0eV, 5.4eV and 4.6eV, representing transition dominated by d → 2π* for 6.0eV, $^1A_1$ → $^5T_2$ for 5.36eV close to the 5.4eV value, representing a d → π* transition. From 3.36eV to 3.94eV, the transition spectrum is dominated by d → π*, ($A_1$, E, $T_1$, $T_2$) to ($T_1$, $T_2$). The absorption cross-sections value for Ni(CO)$_4$ is in the range of ~5.01 x $10^{-17}$cm$^2$.

For Ni(CO)$_4$, the dissociation process follows the reaction equation:

$$Ni(CO)_4 + e- \rightarrow Ni(CO)_4^* \rightarrow Ni(CO)^-_x + n(CO), \quad n = 4-x \qquad (1.11)$$

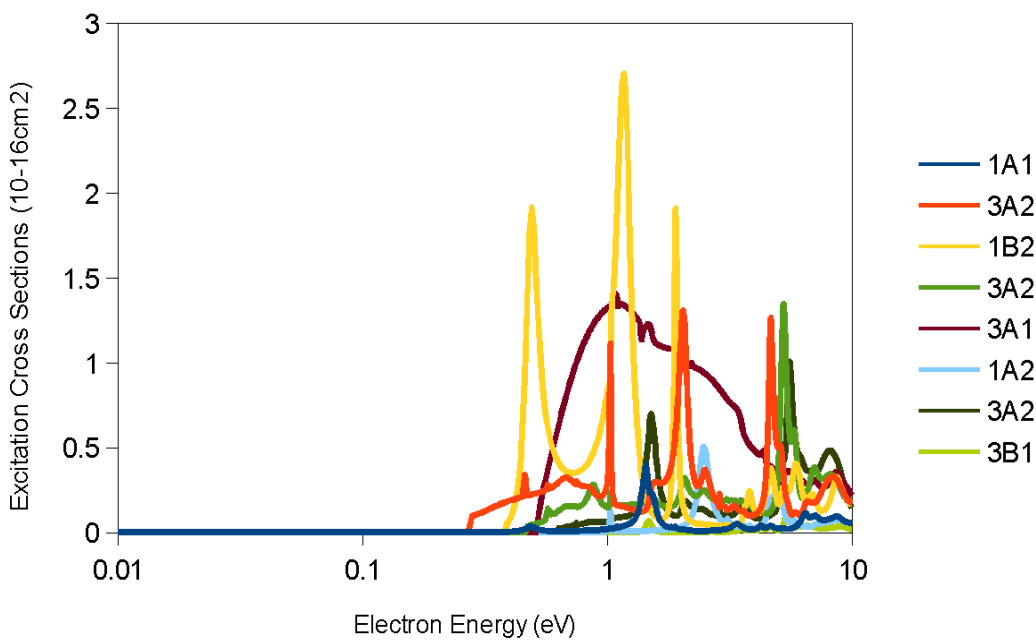

Fig 9. Excitation cross sections of Ni(CO)$_4$

The Ni(CO)$_4$ it is known to strip all four ligands in the process of electron-molecule collision, specific to DEA, so we expect to see all the resonances specific to the molecule undergoing full dissociation, with the negative ions, Ni(CO)$_3^-$, Ni(CO)$_2^-$, NiCO$^-$ and Ni$^-$, at energies between 0.5 – 6eV. As one electron hits one molecule in ground state d and excites it to a higher excited state π* followed by fragmentation, undergoing a allowed transition, the kinetic energy of the molecule increases from E$_G$ to E$_G$' to drop further to E$_{KER}$' and E$_{KER}$" > E$_G$, as we take the fragmentation process as a step by step process. The active space configuration of the theoretical model from our Quantemol calculations is 10A$_g$, 11A$_g$, 12A$_g$, 10A$_u$, 11A$_u$, 10B$_u$, 11B$_u$, 12B$_u$, 11B$_g$.

The molecular structure with symmetry in Cartesian coordinates (X, Y, Z) used for the cross section simulation parameters is presented in ANNEX 1. As well, the bond distances between Ni - C and C - O have been reviewed from multiple sources, in [42] has the values of 1.669Å for Ni – C and 1.153Å for C – O. The bond distances determined in [20] experimentally and through calculations have the value of 1.838Å for Ni – C and 1.142Å for C – O from experiment [43] and 1.831Å for Ni – C and 1.147Å for C – O determined by CCSD calculations using cc - pVTZ basis set. In [46] the Ni - C and C - O bond lengths have a value of 1.84Å and 1.15Å from X-ray single crystal study, and 1.835Å and respectively 1.139Å from gas-phase electron diffraction.

A C$_{2h}$ geometry was employed trying to reduce the calculations steps and the number of iterations. Our Quantemol-N simulations are using a user defined basis set, based on cc-pVTZ and STO-6G. The basis set was user defined to reduce the size of the data and the memory necessary for the simulations.

The values of electron affinity of Ni(CO)$_4$ used for simulation are presented in Table 6.

| Negative Ion | Incident Electron Energy (eV) [41] | Appearance Potentials (eV) [40] | Electron Affinity (eV) [40] | Vibrational Frequency (cm$^{-1}$) [40] |
|---|---|---|---|---|
| Ni(CO)$_3^-$ | 0.8 | 0 | 0.804 ± 0.012 | 2100 ± 80 |
| Ni(CO)$_2^-$ | 1.7 | 1.0 ± 0.4 | 0.643 ± 0.014 | 2100 ± 80 |
| Ni(CO)$^-$ | 4.6 | 3.2 ± 0.5 | 1.077 ± 0.013 | 1940 ± 80 |
| Ni$^-$ | 5.4 | 4.1 ± 0.3 | 1.157 ± 0.010 | |

Table 6. Negative ions of Ni(CO)$_4$ with electron affinity [40]

The DEA cross section from Quantemol-N simulations, presented in Fig. 5, have average values of maximum cross section of 1 – 2 x 10$^{-3}$Å$^2$ at 0.8eV corresponding to the Ni(CO)$_3^-$ fragment, representing the curves with electron affinity values between 1.2 to 1.4eV.

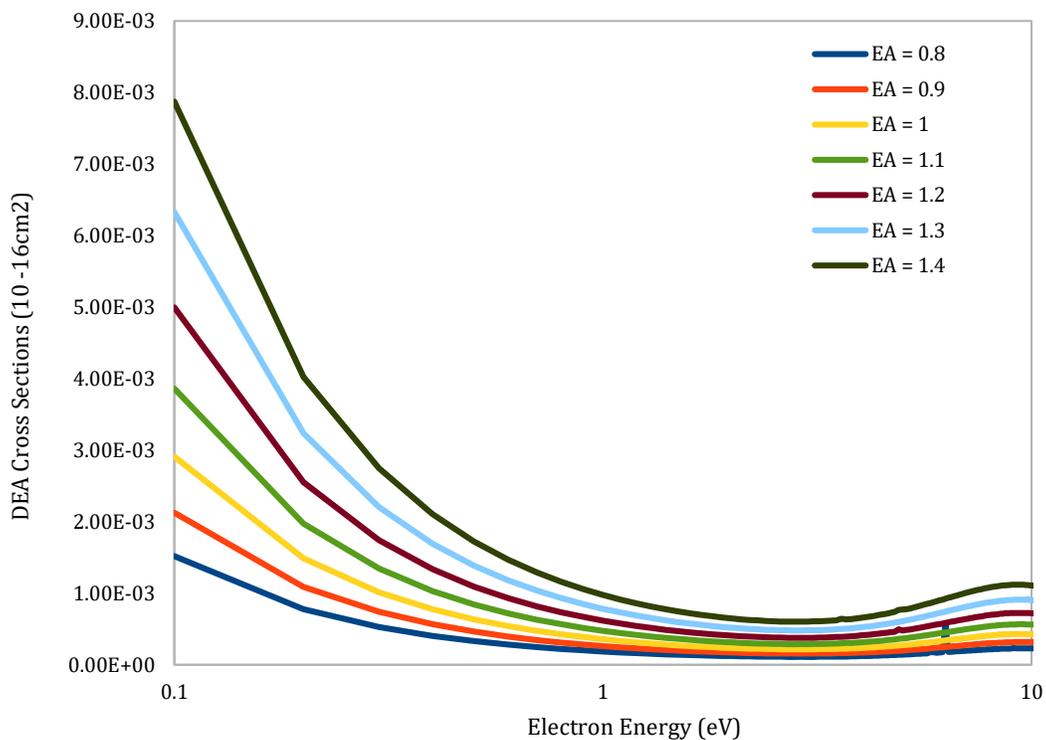

Fig 10. DEA cross sections of Ni(CO)$_4$

Comparatively, taking into account the fragmentation pattern, the violet (color) spectrum has better accuracy and it is closer to the true value, though all present reliable data within the error limit threshold according to the EA(eV) and vibrational frequency used.

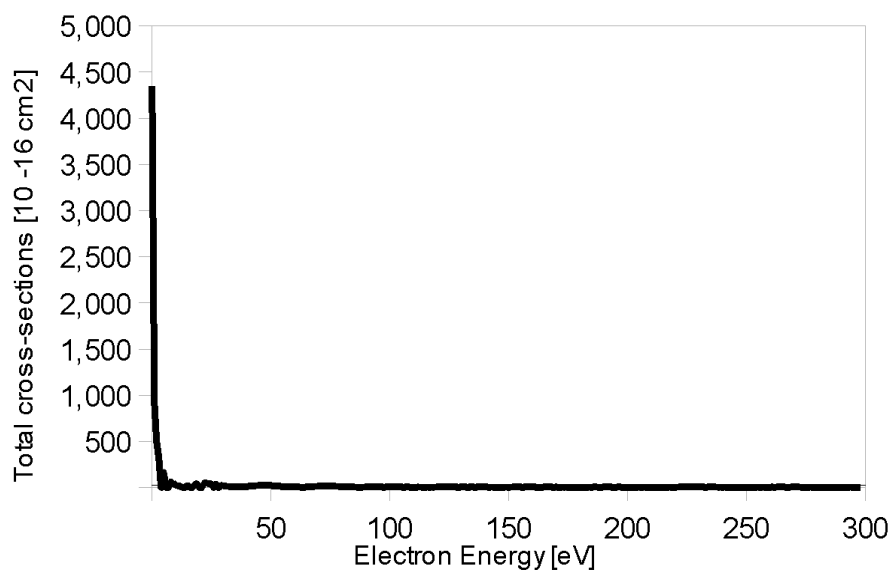

Fig 11. Total cross-sections values for Ni(CO)$_4$, 1eV to 300eV

With the increase in the electron affinity we have an increase in the cross-section values as seen from Fig. 6 and according to that the viability of the best result. The peaks on the curves with lower electron affinity make the result undesirable, and we could argue that the error comes from applying to the simulation a lower value and not from a real resonance.

The total cross-section from Quantemol-N simulations is in close agreement with the reported literature data for ~100-300eV of $2 \times 10^{-16} cm^2$ [47], [48], with a value of $1.6 - 1.8 \times 10^{-16} cm^2$ in the same energy range. [48] makes reference to the same cross-section value of $2 \times 10^{-16} cm^2$ in the energy range of 100-300eV and with the growth of $Ni_x(CO)_y$ species on Ag(111) and carbonyl groups spontaneous desorption between 200-400K resulting pure Ni on the substrate. The total cross-section value is presented in Fig. 8.

**Conclusions**

With increasing importance to cancer research, materials and superconductors development, nanotechnology and focused electron beam deposition and the scarcity of available data on cross sections, in particular for metal containing compounds, the present article aims to bring an insight into the values of dissociative attachment cross sections obtained through Quantemol-N simulations of the three of the most commonly used compounds, commercially available: $Ni(CO)_4$, $Co(CO)_3NO$ and $Cr(CO)_6$.

As the use of Quantemol-N simulations would not necessarily replace experimental data, the results show promising use of the software for reliable cross section data that can be utilized in the multitude of the industrial processes requiring an accurate value of these. The approach we employed, based on R-matrix calculations is rather a simple method making use of basic molecular data, structure and symmetry of the molecule, bonds lengths and specificity, all available and easy to obtain from literature or experimental data.

ANNEX 1:

Atomic structure and X, Y, Z configuration of Cr(CO)$_6$:

| Atom Label | X [Å] | Y [Å] | Z [Å] |
|---|---|---|---|
| C1 | 1.34 | 1.34 | 0 |
| C2 | -1.34 | -1.34 | 0 |
| C3 | 0 | 0 | 1.9 |
| C4 | -1.34 | 1.34 | 0 |
| C5 | 0 | 0 | 0 |
| C6 | 2.15 | 2.15 | 0 |

| Cr7 | -2.15 | -2.15 | 0 |
| O8 | 0 | 0 | 3.04 |
| O9 | -2.15 | 2.15 | 0 |
| O10 | 0 | 0 | -3.04 |
| O11 | 2.15 | -2.15 | 0 |
| O12 | 0 | 0 | -3.04 |
| O13 | 0 | -3.04 | 0 |

Atomic structure and X, Y, Z configuration of Co(CO)$_3$NO:

| Atom Label | X [Å] | Y [Å] | Z [Å] |
|---|---|---|---|
| Co1 | -0.1 | 0 | 0 |
| C2 | 0.66 | -0.81 | -1.4 |
| C3 | 0.66 | -0.81 | 1.4 |
| C4 | 0.66 | 1.62 | 0 |
| O5 | 1.12 | -1.34 | 2.32 |
| O6 | 1.12 | -1.34 | -2.32 |
| O7 | 1.11 | 2.68 | 0 |
| O8 | -2.92 | 0 | 0 |
| N9 | -1.76 | 0 | 0 |

Atomic structure and X, Y, Z configuration of Ni(CO)$_4$:

| Atom Label | X [Å] | Y [Å] | Z [Å] |
|---|---|---|---|
| Ni1 | 0 | 0 | 0 |
| C2 | -0.09 | -1.8 | 0.18 |
| C3 | 1.73 | 0.52 | 0.04 |
| C4 | -0.74 | 0.48 | -1.58 |
| C5 | -0.9 | 0.79 | 1.35 |
| O6 | -0.15 | -2.94 | 0.3 |
| O7 | 2.83 | 0.85 | 0.07 |
| O8 | -1.2 | 0.79 | -2.58 |
| O9 | -1.47 | 1.29 | 2.21 |